\documentclass[12pt]{article}\usepackage{graphics}
\begin{document}
\title{Trouble for MAC}
\author{\normalsize B. Holdom and F. S. Roux\\\small {\em Department of
Physics, University of Toronto}\\\small {\em Toronto, Ontario,} M5S1A7,
CANADA}\date{}\maketitle
\begin{picture}(0,0)(0,0)
\put(310,205){UTPT-98-06}
\put(310,190){hep-ph/9804311}
\end{picture}

\begin{abstract} We show that the next-to-leading corrections to the kernel of
the gap equation can be large and of opposite sign to the lowest order kernel, in
the presence of a gauge boson mass. This calls into question the reliability of
the Most Attractive Channel hypothesis.
\end{abstract}
\baselineskip 19pt

There are few rigorous results for the case of dynamical symmetry breaking in
theories which are neither supersymmetric or QCD-like. Efforts to build
realistic models based on this large class of theories have then relied on various
dynamical assumptions. One dynamical assumption in particular has had a
major impact on model building. This is the most attractive channel (MAC)
hypothesis
\cite{a1}, which assumes that symmetry breaking will be dictated by a fermion
bilinear condensing in the channel which is most attractive under the exchange
of one gauge boson. This is a simple prescription which receives support from
QCD and which can be easily applied to other theories. It ties in closely with
the ladder approximation to the gap equation, which is the basis of many studies
of chiral symmetry breaking. 

The gap equation analysis does in fact provide some support for the MAC
hypothesis, in the sense that it provides an estimate of the critical coupling
required for symmetry breaking to occur. This critical coupling is often
significantly less \cite{aa} than what might be expected for a truly strong gauge
coupling, with the latter being \(\alpha \approx 4\pi \). For a critical coupling
smaller than this, it is then somewhat plausible that corrections beyond the
ladder graphs could be rather small. This possibility was reinforced by the
study in
\cite{a}, where corrections to the kernel of the gap equation were considered to
second (next-to-leading) order. The corrections were found to be at the 20\%
level or less.

In \cite{a} the case of an unbroken strong gauge interaction with small \(\beta
\)-function (walking theory) was considered, and the second order kernel was
explored in a certain momentum region (one momentum much larger than
another) in order to obtain analytical results. We will complement that study by
instead considering the second order kernel for the momentum region expected
to dominate in the loop integrations in the effective action. Our main object is
to consider the implications of a possible gauge boson mass, associated with the
breakdown of the strong gauge interaction. The effect of a gauge boson mass in
the fermion gap equation should be considered for consistency, since the MAC
hypothesis claims to differentiate between different symmetry breaking
patterns, some of which cause the gauge symmetry itself to break. We find that
a gauge boson mass can cause the order \(\alpha ^{2}\) term in the kernel to be
as large or larger than the order \(\alpha \) term, and of opposite sign. This
makes the use of the MAC prescription to decide which channel actually does
condense very uncertain.

We will consider \({n_{f}}\) flavors of fermions all in the fundamental
representation of the gauge group \(\mathit{SU}(N)\). The effective action in
euclidean space is
\begin{equation}\Gamma (S)= - \mathrm{Tr}(\ln(S^{{ - 1}})) +
\mathrm{Tr}([S^{{ - 1}} - {\partial}{\!\!\!/}]S) - \mathrm{(2PI\
diagrams)}\end{equation} where the fermion propagator is
\begin{equation}\mathit{S}(p)=(\mathit{Z}(p)[\mathit{p\!\!\!/} + \Sigma
(p)])^{{ - 1}}.\end{equation} Flavor and color indices are implicit. 

It is typical that the dominant contributions to this effective potential come
from momentum scales larger than \(\Sigma (p)\), in which case we can
perform an expansion in powers of \(\Sigma (p)/p\). This can be rigorously
justified \cite{a,b} in the case of a walking theory. A different situation in
which this expansion may be justified will emerge below. If we expand the
effective potential to second order in \(\Sigma (p)\), we find that the leading
piece (zeroth order in \(\Sigma (p)/p\)) of the equation
\begin{equation}\frac {\delta \Gamma (S)}{\delta
\mathit{Z}(p)}=0\end{equation} reads \cite{b}
\begin{equation}(\mathit{Z}(p) - 1)\mathit{p\!\!
\!/}=\frac {\delta {\mathrm{(2PI\ diagrams)}_{S^{0}}}}{\delta S
^{0}(p)}{.\label{a}}\end{equation} where \(S^{0}(p)\) is the massless
fermion propagator.

By making use of (\ref{a}) \cite{b}, the piece quadratic in \(\Sigma (p)\) in the
effective action becomes
\begin{equation}\Gamma (\Sigma ^{2})=\frac {{n_{f}}N}{4\pi ^{2}}\left(\!
\int _{0}^{\infty }dpp\Sigma (p)^{2} - 
\frac {1}{2}\int _{0}^{\infty }\int _{0}^{\infty }dpdk\Sigma (p)
\Sigma (k)\mathit{F}(p, k)\!\right) {\label{b}}\end{equation}
\begin{eqnarray}&&\mathit{F}(p, k)=\frac {pk}{2\pi ^{2}{n_{f}}N
\mathit{Z}(p)\mathit{Z}(k)}\int _{0}^{\pi }d\theta \sin(\theta
)^{2}{d_{\mathit{ij}}}{d_{\mathit{kl}}}{K_{\mathit{ijkl}}}(p, k)
{\label{e}}\\&&{K_{\mathit{ijkl}}}(p, k)=\frac {\delta {\mathrm{(2PI\
diagrams)}_{S^{0}}}}{\delta {S _{\mathit{ij}}^0}(p)\delta
{S_{\mathit{kl}}^0}(k)}{\label{c}}\end{eqnarray}
 \(K\) is a truncated 4-point function which incorporates the appropriate
symmetry factors of the original 2PI diagrams. The \(d\)'s are constant matrices
in color/flavor/spinor space and are nontrivial only in color space, where they
reflect the representation \(R\) of \(\mathit{SU}(N)\) in which the fermion
mass lies.

When the momentum integrals in (\ref{b}) are defined with an infrared cutoff
\(\kappa \), for example \(\kappa \approx \Sigma (\kappa )\), then \(\Sigma (p)\)
is determined nonperturbatively from the gap equation
\begin{equation}\frac {\delta \Gamma (S)}{\delta \Sigma
(p)}=0.\end{equation}
 \(Z\) and \(K\) are determined perturbatively from (\ref{a}) and (\ref{c}),
where massless fermion propagators are used. We renormalize in the
\(\overline{\mathrm{MS}}\) scheme, so that \(Z\) and \(K\) become functions
of the renormalization scale \(\mu \), the gauge coupling \(\alpha (\mu )\) and
the gauge parameter \(\xi (\mu )\). We will choose Feynman gauge \(\xi (\mu
)=1\). The usual analysis we are exploring displays a gauge dependence, as
explicitly obtained in \cite{a}. This due to the presence of a nonlocal fermion
mass and only a local gauge boson-fermion vertex in the effective action.
Gauge independence would require additional contributions to gauge
boson-fermion vertices proportional to the nonlocal mass. In the case of a
massive gauge boson there is a Goldstone boson-fermion vertex which is also
proportional to the nonlocal fermion mass. We shall restrict ourselves to the
usual analysis and not consider any diagram which introduces a dependence on
the fermion mass through a vertex.

We shall investigate the perturbative expansion of the kernel
\begin{equation}\mathit{F}(p, k)=\frac {2}{\pi }({C_{2}}({r_{1}}) +
{C_{2}}({r_{2}}) - {C_{2}}(R))({F_{1}}(p,  k)\alpha  +
{\cal{R}}{F_{2}}(p, k)\alpha ^{2}) +  {...}{\label{pk}}\end{equation}
Notice that the well known group theory factor appearing in the first order term
also multiplies the second order term. This factor includes all the dependence
on the representation \(R\) of the condensing channel (\({r_{1}}\) and
\({r_{2}}\) are the representations of the two fermions). The appearance of this
at second order is a result of keeping only the diagrams which are leading in
either \(N\) or \({n_{f}}\), as displayed in Fig.\ (1).\footnote{When \(R\) is the
singlet channel, the nonleading diagrams are suppressed by \(1/N^{2}\);
otherwise the suppression is \(1/N\). And we note in particular that the
nonleading crossed-ladder graph is small even without the group theory
suppression.}

At first order
\begin{eqnarray}&&{F_{1}}(p, k)=(p^{2} + k^{2} + M^{2} - \sqrt{(p^{2} +
k^{2} +  M^{2})^{2} - 4p^{2}k^{2}}){/}(2pk)\\&&{\ \ \ \ \ \ \ \ \ \
}=\min(\frac {p}{k}, \frac {k}{p})
\mathrm{\ \ for\ \ }M=0\end{eqnarray} where \(M\) is a possible gauge boson
mass. The \({\cal{R}}\) in (\ref{pk}) will be defined by normalizing
\({F_{2}}(p, k)\) relative to \({F_{1}}(p, k)\) at some momenta which
dominates the integrations in the effective action (\ref{b}), as follows.

The integrals over \(p\) and \(k\) in (\ref{b}) could be exchanged for integrals
over \(\sqrt{(p^{2} + k^{2})/2}\) and \(p/k\). There is then some momentum
\({q_{\mathrm{dom}}}\) which dominates the integral over \(\sqrt{(p^{2} +
k^{2})/2}\).  This scale is determined dynamically by the form of \(\Sigma
(p)\), and it makes sense to set the renormalization scale \(\mu
={q_{\mathrm{dom}}}\). Note that for an unbroken gauge symmetry in the
walking limit, \({q_{\mathrm{dom}}}\) is much larger than \(\Sigma (\kappa
)\) \cite{b}. There is another scale in the problem when the gauge boson has
mass \(M\). We do not expect that \({q_{\mathrm{dom}}}\) will be much
smaller than \(M\), since \(M\) acts as a natural infrared cutoff.  On the other
hand if \({q_{\mathrm{dom}}}\) is much larger than \(M\) then we revert to
the previous case in which the gauge boson mass plays no role in the dynamics.
Thus in the massive case we assume that the dynamics is such that
\({q_{\mathrm{dom}}}\) and \(M\) are of the same order, and so we set
\({q_{\mathrm{dom}}}=M=\mu \).

We are therefore led to the following normalization of \({F_{2}}(p, k)\) at the
point \(p=k={q_{\mathrm{dom}}}\), where we note that the maximum value
of \({F_{1}}\) occurs for \(p=k\) for a fixed \(p^{2} + k^{2}\).
\begin{eqnarray}&&{F_{2}}={F_{1}}\mathrm{\ \ for\ \ }p=k=\mu
=M\\&&{F_{2}}={F_{1}}=1\mathrm{\ \ for\ \ }p=k=\mu 
\mathrm{\ \ when\ \ }M=0\end{eqnarray} With this normalization, the size of
\({\cal{R}}\) provides a measure of the importance of higher order effects. It is
a conservative estimate, since the \(\alpha \) in (\ref{pk}) could be larger than
unity and closer to \(4\pi \) if the interaction is truly strong. Our measure of the
higher order effects differs somewhat from that in \cite{a}; there the
simplifying assumption \({F_{2}}(p, k)\propto \min(\frac {p}{k}, 
\frac {k}{p})\) was made, in which case it was sufficient to evaluate the kernel
at the point \(p\gg k\).

To obtain \({\cal{R}}\) we perform the angular integration in (\ref{e})
numerically. When the gauge boson is massless we find
\begin{equation}{{\cal{R}}_{\mathrm{massless}}}=.33{C_{2}}(G) -
.21{n_{f}}\mathit{T}(r) + {C_{2}}(r)/2\pi {.
\label{k}}\end{equation} Here \(\mathit{T}(r)\equiv (\mathit{T}({r_{1}}) +
\mathit{T}({r _{2}}))/2\) and \({C_{2}}(r)\equiv ({C_{2}}({r_{1}}) +
{C_{2}}({r_{2}})) /2\). The last term arises from the \(Z\) factors in (\ref{e}),
the second term from the fermion loop diagram, and the first term from the
other diagrams in Fig.\ (1). Thus when the number of fermions is small, the
second order kernel reinforces the attraction found at lowest order in the singlet
channel. For \(N=3\) and \({n_{f}}=3\), with fermions in the fundamental
representation, we have \({{\cal{R}}_{\mathrm{massless}}}=.9\). For
increasing numbers of fermions, \({{\cal{R}}_{\mathrm{massless}}}\)
decreases, and it becomes negative when the number of fermions are large
enough to result in a small \(\beta \)-function, corresponding to the condition
\(11N=4{n_{f}}\mathit{T}(R)\). For example for \(N=3\) and \({n_{f}}=16\)
we have \({{\cal{R}}_{\mathrm{massless}}}= - .4\). We find that the
dependence of \({F_{2}}(p, k)\) on \(p/k\) for fixed \(p^{2} + k^{2}\) is quite
different from that displayed by \({F_{1}}(p, k)\). This is the main source of
difference between our result and that in \cite{a}.

We now consider the case that the gauge boson is massive. We will continue to
use Feynman gauge and will ignore the additional diagrams involving
Goldstone bosons which are all subleading in \(1/N\). To be specific we
consider the breakdown of \(\mathit{SU}(N)\) to \(\mathit{SU}(N - 1)\),
corresponding to a fermion mass in the symmetric tensor representation, with
the single massive fermion being a \(\mathit{SU}(N - 1)\) singlet. We find
\begin{equation}{{\cal{R}}_{\mathrm{massive}}}= - .26{C_{2}}(G) -
.045{n_{f}}\mathit{T}(r) - .12{C_{2}}(r).{\label{d}}\end{equation} For
example for \(N=3\) and \({n_{f}}=(3, 16)\) we have
\({{\cal{R}}_{\mathrm{massive}}}=( - 1.0,  - 1.3)\). Compared to the
massless case we also find that \({F_{1}}(p, k)\) and \({F_{2}}(p, k)\) have a
much more similar dependence on \(p/k\) for fixed \(p^{2} + k^{2}\). Thus we
conclude that the order \(\alpha ^{2}\) term in the kernel of the gap equation is
large and of opposite sign to the order \(\alpha \) term. 

The main uncertainty in \({{\cal{R}}_{\mathrm{massive}}}\) is that the true
values of \({q_{\mathrm{dom}}}\), \(M\), and \(\mu \) may in fact deviate
from our assumption of equality. We stress though that
\({q_{\mathrm{dom}}}\) is dynamically determined from the form of \(\Sigma
(p)\), and its value should be such as to maximize the attraction in the preferred
channel. A more minor source of uncertainty is in the relation of the two gauge
boson masses corresponding to diagonal and nondiagonal generators
respectively (we have denoted the latter mass by \(M\)). The numbers
appearing in (\ref{d}) assume that the former is \(\sqrt{\frac {4}{3}}\) times as
large as the latter, as may be expected when \(N=3\).

It appears that the possibility of a symmetry breaking solution is
self-consistent, in the sense that attraction in this channel is a consequence of
the gauge boson mass which in turn is consistent with the symmetry breaking
fermion mass. On the other hand other conditions must be satisfied before
gauge symmetry breaking can occur; in particular the theory cannot be purely
vectorlike
\cite{d}. Besides theories which are explicitly chiral, additional gauge
interactions or nonrenormalizable interactions generated by physics at a higher
scale may make a theory effectively chiral. Our analysis also applies to
situations where the source of the gauge boson mass is something other than the
strong dynamics in question.

We now return to the validity of the expansion in \(\Sigma (p)/p\), which has
been justified so far only for the case of an unbroken, walking theory. The
expansion would be justified in the broken theory if a fermion mass smaller
than \(M\) emerged, since \(M\) sets the scale for the dominant momenta in the
loops. We now see that there could be a dynamical reason for this to occur. 
The point is that from (\ref{d}) we see that the fermion loop (the term with
\({n_{f}}\)) tends to enhance the strength of the second order kernel. We
therefore see that a fermion mass small compared to \(M\) enhances the chance
that the mass forms, since a large fermion mass damps the fermion loop
contribution. This is in contrast to QCD where a large quark mass and the
resulting damping of the fermion loop enhances the lowest order attraction in
the color singlet channel (see (\ref{k})). Of course we are only pointing out a
possible mechanism, since higher order effects are important and unknown.

We have explored the case of a broken gauge symmetry and have found that
there is little reason to believe the lowest order most attractive channel
hypothesis. Although this result runs counter to conventional wisdom
concerning MAC, it is not terribly surprising---when the coupling is large,
higher order effects can be important. The problem of gauge dependence also
plagues the usual analysis, but there is no reason to expect that additional
contributions present in a gauge invariant treatment would cancel those that we
have found. Our results are sufficient to call into question the use of the MAC
hypothesis as a test of whether or not gauge symmetries break. More powerful
techniques are needed to study the symmetry breaking patterns of interest for
the construction of realistic theories of mass and flavor.

\section*{Acknowledgment} This research was supported in part by the
Natural Sciences and Engineering Research Council of Canada. BH thanks the
hospitality and support of the KEK theory group, where this work was
completed.

\newpage
\begin{center} \includegraphics{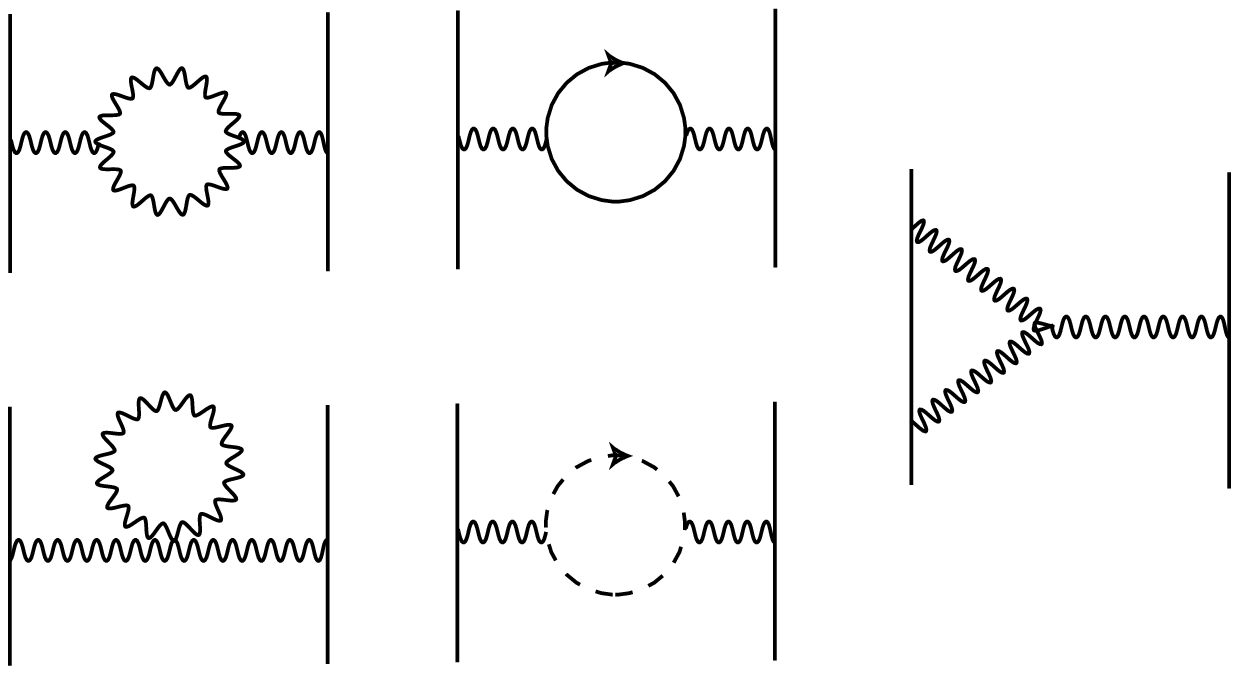}
\end{center}

\noindent Figure (1): The diagrams leading in \(N\) or \({n_{f}}\) which
contribute to the second order kernel.


\begin{thebibliography}{99}
\bibitem{a1} S. Raby, S. Dimopoulos, and L. Susskind, Nucl. Phys.
\textbf{B169}, 373 (1980).
\bibitem{aa} B. Holdom, Phys. Rev. \textbf{D56} (1997) 7461.
\bibitem{a} T. Appelquist, K. Lane and U. Mahanta, Phys. Rev. Lett.
\textbf{61} (1988) 1553.
\bibitem{b} B. Holdom, Phys. Lett. \textbf{B213} (1988) 365; Phys. Rev. Lett.
\textbf{62} (1989) 997.
\bibitem{d} C. Vafa and E. Witten, Nucl. Phys.
\textbf{B234} (1984) 173.
\end{thebibliography}
\end{document}